\documentclass{article}

\usepackage{microtype}
\usepackage{graphicx}
\usepackage{booktabs} 

\usepackage{hyperref}

\usepackage[accepted]{icml2024-preprint}

\usepackage{amsmath}
\usepackage{amssymb}
\usepackage{mathtools}
\usepackage{amsthm}

\usepackage{amsmath,amsfonts,bm}

\def\eqref#1{equation~\ref{#1}}

\def\1{\bm{1}}

\def\ve{{\bm{e}}}

\def\vh{{\bm{h}}}

\def\vp{{\bm{p}}}

\DeclareMathAlphabet{\mathsfit}{\encodingdefault}{\sfdefault}{m}{sl}
\SetMathAlphabet{\mathsfit}{bold}{\encodingdefault}{\sfdefault}{bx}{n}

\def\gB{{\mathcal{B}}}
\def\gC{{\mathcal{C}}}

\def\gF{{\mathcal{F}}}

\def\gH{{\mathcal{H}}}

\def\gR{{\mathcal{R}}}

\def\gX{{\mathcal{X}}}

\def\sI{{\mathbb{I}}}

\usepackage{adjustbox}
\usepackage{multirow}

\usepackage[capitalize,noabbrev]{cleveref}

\theoremstyle{plain}

\theoremstyle{definition}

\theoremstyle{remark}

\usepackage[textsize=tiny]{todonotes}

\begin{document}

\twocolumn[
\icmltitle{Projecting Molecules into Synthesizable Chemical Spaces}



\icmlsetsymbol{equal}{*}
\icmlsetsymbol{corresponding}{\#}

\begin{icmlauthorlist}
\icmlauthor{Shitong Luo}{corresponding,equal}
\icmlauthor{Wenhao Gao}{equal,mit}
\icmlauthor{Zuofan Wu}{equal,helixon}
\icmlauthor{Jian Peng}{helixon}
\icmlauthor{Connor W. Coley}{mit}
\icmlauthor{Jianzhu Ma}{thu}
\end{icmlauthorlist}

\icmlaffiliation{mit}{Massachusetts Institute of Technology}
\icmlaffiliation{helixon}{Helixon Research}
\icmlaffiliation{thu}{Tsinghua University}

\icmlcorrespondingauthor{Shitong Luo}{\mbox{luost26@gmail.com}}

\icmlkeywords{chemical space, synthesizability, molecular generation}

\vskip 0.3in
]



\printAffiliationsAndNotice{} 

\begin{abstract}

Discovering new drug molecules is a pivotal yet challenging process due to the near-infinitely large chemical space and notorious demands on time and resources.
Numerous generative models have recently been introduced to accelerate the drug discovery process, but their progression to experimental validation remains limited, largely due to a lack of consideration for synthetic accessibility in practical settings.
In this work, we introduce ChemProjector, a novel framework that is capable of generating new chemical structures while ensuring synthetic accessibility.
Specifically, we introduce a postfix notation of synthetic pathways to represent molecules in chemical space.
Then, we design a transformer-based model to translate molecular graphs into postfix notations of synthesis.
We highlight the model's ability to: (a) perform bottom-up synthesis planning more accurately, (b) generate structurally similar, synthesizable analogs for unsynthesizable molecules proposed by generative models with their properties preserved, and (c) explore the local synthesizable chemical space around hit molecules. Code is available at \url{https://github.com/luost26/ChemProjector}.

\end{abstract}

\section{Introduction}

Designing novel drug molecules, a process that aims to identify molecular structures with desired properties, is known to be notoriously complex, time- and resource-consuming \citep{wouters2020estimated}.
Methods that could accelerate the discovery process are thus of significant interest in medicinal science and the pharmaceutical industry.
Historically, two principal computational methodologies have been employed in drug design: virtual screening \citep{shoichet2004virtual} and \textit{de novo} design \citep{hartenfeller2011novo}.
Virtual screening involves predicting properties for a predefined set of molecules, known as virtual libraries, and identifying the most promising candidates for subsequent validation.
These libraries are usually commercial catalogs (make-on-demand compounds) from chemical vendors \citep{lyu2019ultra} and generally consist of molecules with assured synthesizability \citep{enamine}.
However, screening large numbers of molecules is a time-intensive process. Moreover, even libraries on the scale of billions present only a tiny fraction of the theoretical chemical space (Figure~\ref{fig:comparison}a), which is nearly infinite in the number of potential synthesizable small molecules \citep{bohacek1996art}.
This means that many potentially valuable molecules might not be included in these libraries. Additionally, as the size of the library increases, so does the cost of screening, which in turn limits the effectiveness and efficiency of virtual screening in the drug discovery process.

\textit{De novo} molecular design methods, unlike virtual screening, dynamically create molecules tailored to specific objectives like protein binding and bioactivity. Early methods of this kind typically adopt combinatorial optimization \citep{venkatasubramanian1995evolutionary,jensen2019graph,brown2019guacamol}, while recently, neural network enhanced optimization \citep{you2018graph,fu2021differentiable} and generative modeling \citep{gomez2018automatic,jin2018junction} have become popular. These methods aren't constrained by the limitations of virtual libraries, enabling the efficient exploration of novel chemical entities (see Figure~\ref{fig:comparison}c). However, a notable defect with most existing generative models is their tendency to propose synthetically infeasible molecules (Figure~\ref{fig:comparison}c) \citep{gao2020synthesizability}, a concern particularly pronounced in structure-based drug design, an area of increasing interest \citep{bohacek1996art,luo20213sbdd,fu2022reinforced}. 
This issue stems from the inherent ignorance of synthetic accessibility in formulations of most existing generative algorithms, in which molecules are constructed using SMILES strings or atoms and motifs as fundamental units \citep{gao2022sample}. 
This often leads to chemical structures that fall outside synthesizable chemical spaces, posing practical challenges in the experimental validation as well as production of such designs.

Incorporating synthesizability as an additional optimization criterion is a common strategy, yet it introduces complex challenges. Synthesizability is difficult to define precisely due to its sensitivity to subtle structural differences, reaction selectivity, and building block availability inherent in chemical reactions \citep{gao2020synthesizability}. While notable efforts have been made to estimate synthetic accessibility \citep{ertl2009estimation,thakkar2021retrosynthetic,vorvsilak2020syba}, reliable quantification of synthetic accessibility is still a far goal. A more effective strategy involves constraining the design process to only consider molecules that are synthesizable, ensuring that algorithms focus exclusively on viable structures. Based on this idea, several models have been introduced \citep{bradshaw2019moleculechef,bradshaw2020barking,gao2021synnet,swanson2023generative}, though these models have notable limitations, such as poor coverage of synthesizable chemical space or inefficiency in optimizing the properties, as further discussed in Section~\ref{sec:related:design}.

To address the recognized challenges, we introduce a framework that learns to ``project'' molecules into synthesizable chemical spaces (Figure~\ref{fig:comparison}d). Given molecules without synthesizability guarantee, our model is able to ``fix'' them by identifying synthesizable analogs that are structurally similar, thus preserving the structural factors that contributed to the desired properties \citep{janela2022simple}.
We represent molecules by a scalable linear representation for synthetic pathways based on postfix notations.
By generating synthetic pathways in the form of postfix notations rather than molecular graphs or strings, we can ensure the designed molecules are derivable from purchasable chemical building blocks and expert-defined chemical reaction rules, thereby guaranteeing the synthesizability of resulting molecules.
We adopt the transformer architecture \citep{vaswani2017attention}, with an encoder for molecular graphs and a decoder to generate postfix notations of synthesis.

We validate the effectiveness of the model by showcasing its ability to generate synthesizable analogs that are structurally and functionally similar to input unsynthesizable molecules.
Furthermore, the model demonstrates its proficiency in navigating the local synthesizable chemical space, proving invaluable for the expansion and exploration of potential hit molecules. The main contributions of our work can be summarized as follows:
\begin{itemize}
  \setlength{\itemsep}{0pt}
  \item We propose a novel linear representation for synthesis pathways based on postfix notations.
  This also provides a uniform and generalizable representation for all experimental procedures.
  \item We present a novel and robust algorithm tailored for synthesizable molecular design. Specifically, our model can project unsynthesizable molecules from existing generative models to synthesizable chemical space to obtain a structural analog with desired properties such as binding and bioactivity preserved, as well as explore local chemical space for hit expansion.
  \item We demonstrate strong empirical performance in evaluations of bottom-up synthesis planning and analog design, outperforming prior methods. 
\end{itemize}

\section{Related Work}
\label{sec:related:design}

\begin{figure}[t]
\begin{center}
\centerline{\includegraphics[width=0.95\columnwidth]{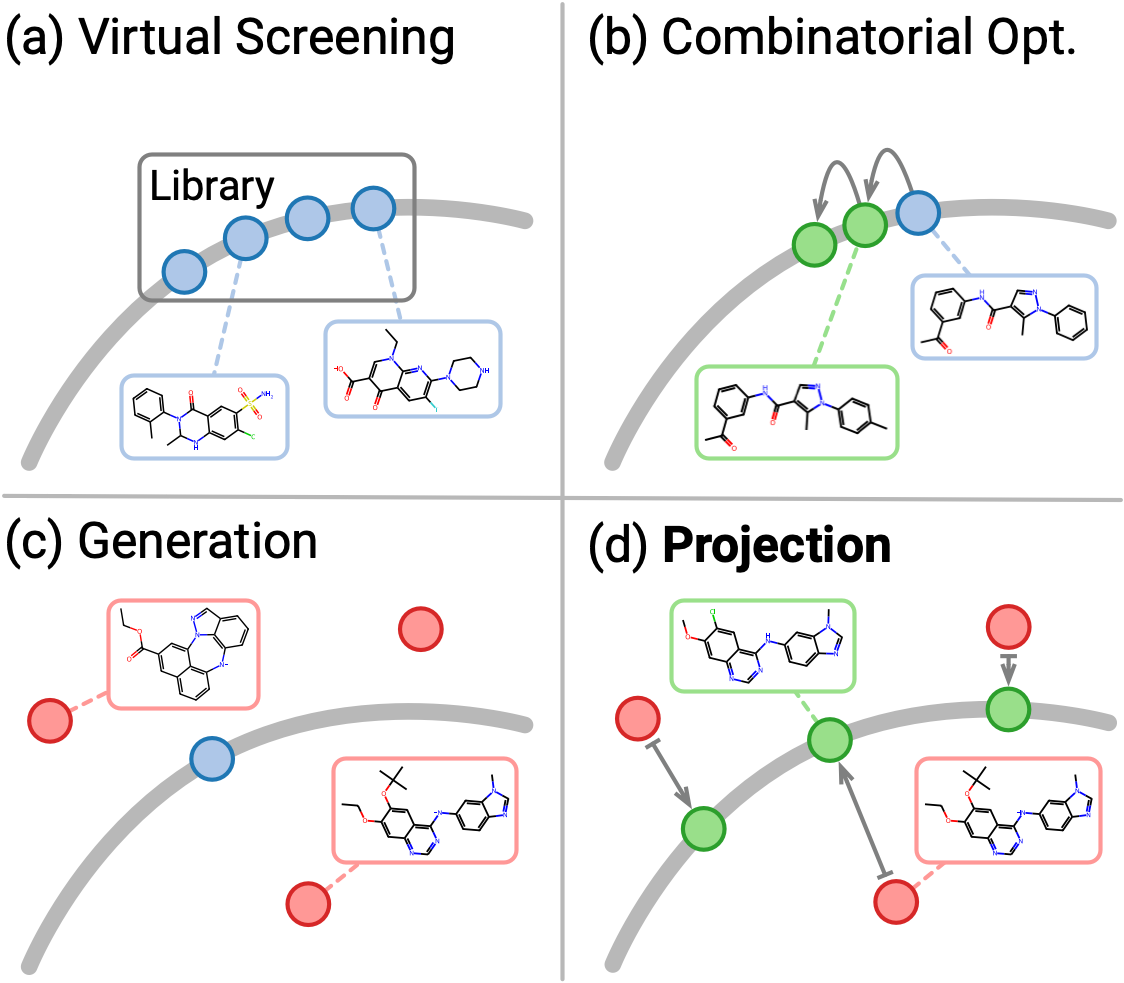}}
\caption{
  Illustration of \textbf{(a)} virtual screening algorithms, \textbf{(b)} combinatorial optimization algorithms, \textbf{(c)} \textit{de novo} generative models, and \textbf{(d)} projection to synthesizable chemical space. Each dot represents a chemical structure, and the gray curve represents a synthesizable chemical space.
}
\label{fig:comparison}
\end{center}
\vspace{-15pt}
\end{figure}

We divide the synthesizable molecular design methods into combinatorial optimization algorithms, reinforcement learning, and autoencoders. We acknowledge and emphasize that the boundaries between these categories are not clear-cut and many intersections exist. This classification is primarily for the sake of clarity in our discussion. 

\textbf{Combinatorial optimization algorithms} use a set of chemical building blocks and reaction templates to make the chemical space combinatorial, and then apply combinatorial optimization algorithms, such as genetic algorithms \citep{gao2021synnet}, Monte Carlo tree search \citep{swanson2023generative}, and Bayesian optimization \citep{korovina2020chembo}, to search for desired molecules (Figure~\ref{fig:comparison}b).
Early representative work before the advent of molecular deep learning includes SYNOPSIS \citep{vinkers2003synopsis} and DOGS \citep{hartenfeller2012dogs}, which perform virtual reactions to generate candidate molecules and select candidates according to scoring functions.
In recent years, advances in deep learning for chemical reactions \cite{coley2019reactivity,schwaller2019molecular} have inspired methods such as MoleculeChef \citep{bradshaw2019moleculechef}, ChemBO \citep{korovina2020chembo}, and DoGs \citep{bradshaw2020barking} that use neural networks to predict reaction outcomes in a ``template-free'' manner.

\textbf{Reinforcement learning approaches}, exemplified by PGFS \citep{gottipati2020pgfs} and REACTOR \citep{horwood2020reactor}, formulate the forward synthesis process as Markov decision processes and train agents to explore the chemical space under the guidance of reward functions that quantify desired properties. 
The agent is provided with a set of reactants and reaction templates and learns to synthesize desirable molecules with them.
As the agent is aware of the objective, its search process is directed and expected to be more efficient.

\textbf{Autoencoder-based methods}, represented by MoleculeChef \citep{bradshaw2019moleculechef} and DoGs \citep{bradshaw2020barking}, leverage the autoencoder architecture to model synthetic pathways in a generative manner.
MoleculeChef is one of the earliest neural models of this kind.
Instead of synthetic paths, it tries to encode and decode a bag of starting reactants from purchasable building blocks. Following an enumeration of the one-step synthetic pathways, the ``best'' molecule from the products is chosen as the output.
DoGs represents synthetic pathways as directed acyclic graphs and serializes the graph construction process as a sequence of actions that includes node addition, reactant selection, and node connection.
DoGs use recurrent neural networks to generate action sequences from latent codes, enabling \textit{de novo} molecular generation by drawing samples from the latent space. SynNet \citep{gao2021synnet} can also be seen as this kind, in which the molecular fingerprinting algorithm is the encoder, and neural networks are used as a decoder to generate synthetic paths.

\textbf{Limitations:} While the synthesizable molecular design methods have advanced significantly, they are still in the proof-of-concept stage and far from mature.
The efficiency of combinatorial optimization often falls short, as these algorithms edit molecules locally, thus being inefficient to explore the vast chemical space. 
Reinforcement learning techniques depend heavily on querying oracle functions, which can hinder their practical application.
MoleculeChef \citep{bradshaw2019moleculechef} primarily explores one-step pathways due to the computational impracticality of enumerating an exponentially growing number of candidates.
Models like PGFS \citep{gottipati2020pgfs} and REACTOR \citep{horwood2020reactor} are limited to generating only linear synthetic pathways. 
In addition, the capacity of these models to generalize and construct complex synthetic paths is yet to be convincingly demonstrated.
Models such as DoGs \citep{bradshaw2020barking} and SynNet \citep{gao2021synnet}, theoretically capable of encompassing a complete chemical space, have not been empirically validated for their ability to adapt to novel, unseen chemicals.
Additionally, these models are generally less adept at generating convergent pathways or more intricate molecular structures, indicating a need for further refinement in their design and functionality.

\begin{figure*}[ht]
\begin{center}
\centerline{\includegraphics[width=0.95\linewidth]{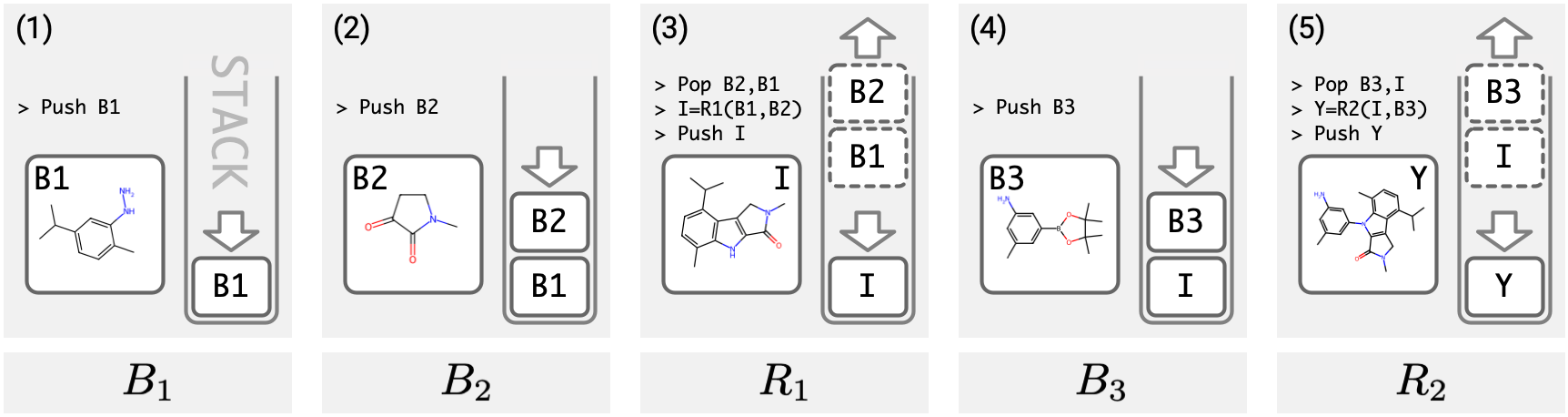}}
\caption{
  \textbf{Top:} Simulation of the synthetic plan $R_2(R_1(B_1, B_2), B_3)$ with a memory stack to store intermediate molecules. In the first two steps, building blocks $B_1$ and $B_2$ are pushed onto the stack.
  In the third step, $B_1$ and $B_2$ are popped from the stack and used as input for reaction $R_1$, leading the intermediate reaction product $I$, which is then added back onto the stack.
  Subsequently, in the fourth step, building block $B_3$ is pushed onto the stack.
  Finally, reaction $R_2$ is applied to $I$ and $B_3$, leading to the final product $Y$.
  \textbf{Bottom:} The postfix notation of the synthetic pathway is $[B_1, B_2, R_1, B_3, R_2]$. Each symbol in the notation corresponds to one step in the synthesis.
}
\label{fig:stack}
\end{center}
\vspace{-15pt}
\end{figure*}

\section{Method}

\subsection{Preliminaries}

\paragraph{Chemical space}
A chemical space $\gC$ is a set of molecules generated by a finite set of \textit{building block} molecules $\gB = \{ B_1, B_2, \ldots B_N \} \subseteq \gC$ and a set of \textit{reaction} rules $\gR = \{ R_1, R_2, \ldots R_M \}$.
Formally, a reaction rule maps reactants to a reaction product. For example, a reaction rule $R$ with two reactants can be denoted by:
\begin{equation}
\label{eq:rxn-rule}
\begin{aligned}
    R: &  \gX_1 \times \gX_2 \rightarrow \gC  \\
       & (X_1, X_2) \mapsto Y,
\end{aligned}
\end{equation}
where $\gX_1$ and $\gX_2$ are sets of molecules to which reaction $R$ can be applied, and $Y$ is the \textit{main} reaction product.
The chemical space is generated by starting from the building blocks and iteratively applying the reactions to every possible combination of molecules.
There are some cases where a reaction has multiple possible \textit{main} products.
For ease of discussion, we assume that reactions produce only one main product in this section, though the actual implementation of our model supports multiple products.

Every molecule in the chemical space is a product of a synthetic process, which involves applying reaction rules to building blocks and intermediate molecules to build a complex molecule. For example, $R_2(R_1(B_1, B_2), B_3)$ denotes a process in which we first apply reaction $R_1$ to building blocks $B_1$ and $B_2$, and then apply reaction $R_2$ to building block $B_3$ and the product of $R_1(B_1, B_2)$.

\paragraph{Postfix notation of synthesis}
We formalize the synthetic process as a sequence of function call instructions with a memory stack to store arguments and intermediate results.
The instruction sequence contains two types of operations: pushing a building block molecule onto the stack and applying a reaction.
The reaction operation first pops the required number of molecules at the top of the stack, then computes the reaction product which will be pushed back onto the stack.
The two operations can be denoted by the building block and the reaction associated with them, leading to the postfix notation of the synthetic process.

Following the above example, the instruction sequence for $R_2(R_1(B_1, B_2), B_3)$ is: (1) push $B_1$; (2) push $B_2$; (3) apply $R_1$; (4) push $B_3$; (5) apply $R_2$, which can be simplified as the postfix notation: $[B_1,B_2,R_1,B_3,R_2]$ (Figure~\ref{fig:stack}).

The synthetic process can be parsed to an abstract syntax tree (AST), which is equivalent to the synthetic tree used in previous work and the postfix notation is a post-order traversal sequence of the AST.
Although it has connections with previous methods, the proposed postfix notation has multiple advantages.
First, the postfix notation is capable of representing branched synthesis pathways, where reactions can be applied to more than one intermediate deposited in the stack. In contrast, previous methods \citep{gottipati2020pgfs,horwood2020reactor} are limited to linear synthetic pathways, where intermediates can only react with building blocks.
In addition, the postfix notation is a simple representation comprising only building blocks and reactions, whereas previous synthetic tree-based representations require extra symbols, such as node addition and edge addition, to construct a synthetic plan.
Lastly, the postfix notation allows reactions with multiple (more than two) molecule inputs, while the previous representation \citep{gao2021synnet} is incompatible with such complex reaction rules.
Therefore, the postfix notation can be immediately applied to chemical spaces that include reactions with reagent specification or with multiple reactants, showing the potential for more complex settings in future work.

\vspace{-10pt}
\paragraph{Problem definition}
We formulate the task of this work, projecting molecules into the chemical space, as generating a postfix notation of synthesis from the input molecular graph, so that the postfix notation denotes a molecule in the chemical space that is identical or structurally similar to the input molecule.

\begin{figure*}[ht]
\begin{center}
\centerline{\includegraphics[width=\linewidth]{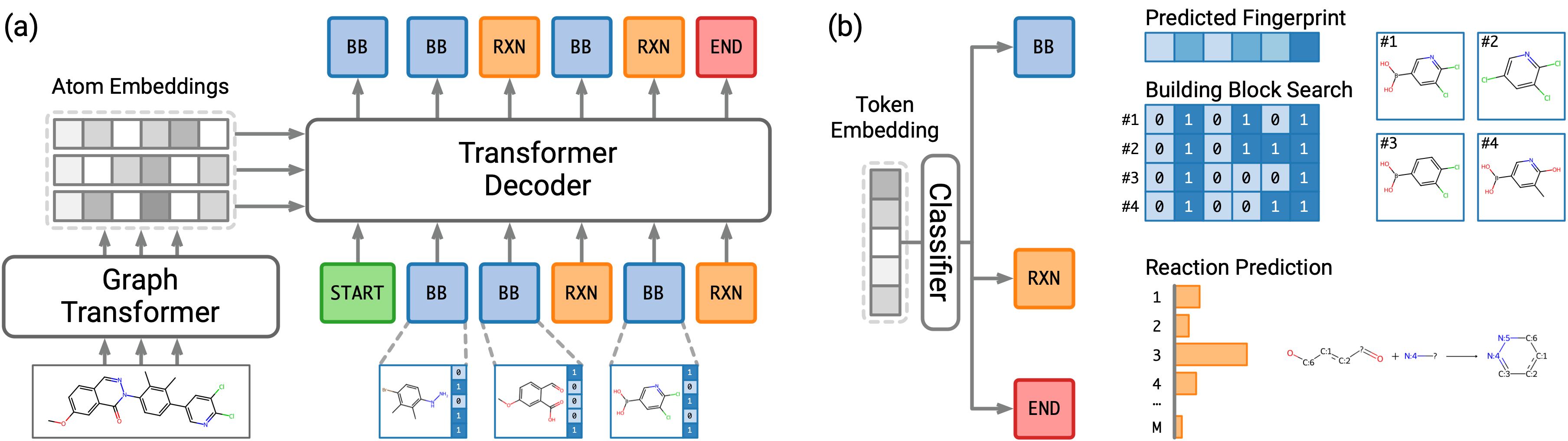}}
\caption{
  \textbf{(a)} Overall architecture of the network. A graph transformer encodes the input molecular graph into atom embeddings. A transformer decoder autoregressively decodes the postfix notation, with atom embeddings as the context.
  \textbf{(b)} Architecture of the prediction network. First, a classifier predicts the token type from the token embedding. If the predicted token type is \texttt{[BB]} (building block), the token embedding is used to predict the building block fingerprint to retrieve building blocks. If the token type is \texttt{[RXN]} (reaction), the token embedding is used to predict the type of reaction. Otherwise, $\texttt{[END]}$ marks the end of the postfix notation.
}
\label{fig:network}
\end{center}
\vspace{-15pt}
\end{figure*}

\subsection{Model}
The model consists of two components: a graph transformer encoder network that encodes the molecular graph and a transformer decoder network that autoregressively generates postfix notations.

\vspace{-5pt}
\paragraph{Molecular graph encoder}
A molecule is represented by a graph with a set of atoms and bonds.
The graph encoder network first converts each atom into initial embeddings according to the atomic number.
Then a stack of graph transformer layers \citep{ying2021graphormer}, which take edges into account by adding a bond type-dependent learnable bias term to the query-key product matrix, performs multi-head self-attention on the atoms.
Finally, the output per-atom embeddings are used for decoding.

\paragraph{Postfix notation decoder}
A postfix notation of synthesis is a sequence that contains four types of tokens: building block tokens, reaction tokens, a start token \texttt{[START]}, and an end token \texttt{[END]}.
A building block token is attached with the fingerprint of the building block molecule, denoted by \texttt{[BB,$\vp$]}, where $\vp \in \{ 0, 1\}^{256}$ is the Morgan fingerprint of length 256 and radius 2 \citep{morgan1965generation}.
Building block tokens are converted into embeddings by an MLP.
A reaction token is the index $r$ of the reaction denoted by \texttt{[RXN,$r$]}. 
The reaction tokens and the start token are converted into embeddings by looking up from the embedding table.
Sinusoidal positional encodings \cite{vaswani2017attention} that indicate the order of tokens are added to the embeddings.
In summary, the initial embedding of the $i$-th token $T_i$ is defined as:
\begin{equation}
    \vh_i^{(0)} = \operatorname{PE}(i) + \begin{cases}
      \ve_\text{start}, & \operatorname{type}(T_i) = \texttt{START} \\
      \operatorname{MLP}_\text{fp}(\vp_i), & \operatorname{type}(T_i) = \texttt{BB} \\
      \operatorname{Embed}_\text{rxn}(r_i), & \operatorname{type}(T_i) = \texttt{RXN}
    \end{cases}.
\end{equation}

Next, a stack of standard transformer layers \cite{vaswani2017attention} takes the sequence embeddings along with the molecular graph embeddings as input, and then predicts the next token using the embeddings from the last layer $\vh_i$ for each position of the sequence.
The prediction network first predicts the type of the next token:
\begin{equation}
    \operatorname{type}(\hat{T}_{i+1}) \sim \operatorname{softmax}( \operatorname{MLP}_\text{type} (\vh_i) ),
\end{equation}
where possible token types include \texttt{BB}, \texttt{RXN}, and \texttt{END}.
If the next token is predicted to be a reaction token \texttt{BB}, the fingerprint will be predicted and the building block will be retrieved via nearest-neighbor search:
\begin{align}
    \hat{\vp}_{i+1} & = \operatorname{sigmoid}(\operatorname{MLP}_\text{fp-pred}(\vh_i)), \\
    \hat{B}_{i+1} & = \operatorname{Nearest-Neighbor}(\hat{\vp}_{i+1}, \{ B_1, \ldots B_N\}),
\end{align}
where $\{ B_1, \ldots B_N\}$ is the building block set of the chemical space. 
The fingerprint of the retrieved $\hat{B}_{i+1}$ is used to replace the predicted fingerprint as input to the model before predicting the next token.
The nearest-neighbor search could produce multiple building block candidates, enabling the generation of diverse pathways.
If the next token is predicted to be a reaction token, the reaction type will be predicted by:
\begin{equation}
    \hat{r}_{i+1} \sim \operatorname{softmax}( \operatorname{MLP}_\text{rxn} (\vh_i) ).
\end{equation}
Similarly, the sampled reaction template $\hat r_{i+1}$ is used as input to the decoder to predict the next token.

\subsection{Training}
We sample synthetic pathways from the chemical space and convert product molecules into graphs to create (postfix notation, graph) pairs for training.
The synthesis sampling algorithm iteratively selects building blocks and applicable reactions at random to generate valid operation sequences as described in Algorithm~\ref{alg:train} in Appendix.
Synthetic pathways are generated on-the-fly at each training step.

We use the standard technique to train transformer decoders in parallel \citep{vaswani2017attention} --- feeding the complete sequence into the decoder, offseting the output embeddings by one position, and applying causal masking.

The training loss consists of three terms corresponding to the three prediction heads.
The first term is a cross entropy loss for token type:
\begin{equation}
    L_\text{type} = \frac{1}{\ell} \sum_{i=0}^{\ell - 1} \operatorname{CE}\left( \operatorname{type}(\hat{T}_{i+1}), \operatorname{type}(T_{i+1}) \right),
\end{equation}
where $\ell$ is the length of the sequence.
The second term is a loss function for building block fingerprints:
\begin{equation}
\begin{aligned}
    L_\text{bb} & = \frac{1}{n} \sum_{i=0}^{\ell-1} \sI\left( \operatorname{type}(T_{i+1}) = \texttt{BB} \right) \cdot \\
    & \qquad \qquad \qquad \sum_{j=0}^{255} \operatorname{BCE}\left( \hat{\vp}_{i+1}[j], \vp_{i+1}[j] \right),
\end{aligned}
\end{equation}
where $\sI$ is the indicator function, BCE is the binary cross entropy loss function, and $n = \sum_{i=0}^{\ell-1} \sI \left( \operatorname{type}(T_{i+1}) = \texttt{BB} \right)$ is the number of \texttt{BB} tokens in the sequence.
The last term is a cross entropy loss for reaction type:
\begin{equation}
    L_\text{rxn} = \frac{1}{m} \sum_{i=0}^{\ell-1} \sI\left( \operatorname{type}(T_{i+1}) = \texttt{RXN} \right) \cdot \operatorname{CE}\left( \hat{r}_{i+1}, r_{i+1} \right).
\end{equation}
The final loss function is the sum of the three terms: $L = L_\text{type} + L_\text{bb} + L_\text{rxn}$.

\begin{table*}[tb]
\vspace{-10pt}
  \caption{
    The proposed method achieves higher success rate, reconstruction rate, and similarity scores in comparison to the baseline model, indicating that our model can find synthetic pathways for synthesizable molecules more accurately.
  }
  \label{tab:generalize}
  \vskip 0.1in
  \centering
  \small
  \begin{adjustbox}{max width=1\textwidth}
  \begin{tabular}{ll ccccc}
\toprule
Dataset & Method & Success\% & Recons.\% & Sim.(Morgan) & Sim.(Scaffold) & Sim.(Gobbi) \\
\midrule
\multirow{2}{*}{Test Set} & SynNet & 84.1\% & 10.7\% & 0.4575 & 0.5109 & 0.3465 \\
& \bf ChemProjector & \bf 97.5\%  & \bf 28.4\% & \bf 0.7167 & \bf 0.7791 & \bf 0.7273 \\

\midrule
\multirow{2}{*}{ChEMBL} & SynNet & 85.0\% & 5.4\% & 0.4270 & 0.4174 & 0.2678 \\
& \bf ChemProjector & \bf 98.8\% & \bf 13.3\% & \bf 0.5978 & \bf 0.5869 & \bf 0.5570 \\

\bottomrule
\end{tabular}

  \end{adjustbox}
\end{table*}

\subsection{Inference}
At inference time, we create an empty stack and generate tokens one by one.
When the generated token is a building block token, we look up the building block molecule according to the predicted fingerprint and push the building block onto the stack.
When the generated token is a reaction token, we first pop molecules from the stack according to the number of required reactants, then use RDKit to predict the product according to the reaction template, and finally push the product onto the stack.
If the number of molecules on the stack is less than the required number or if the reaction is not applicable to the molecules, the inference process will be aborted.
Finally, an end token \texttt{[END]} terminates the process and the molecule at the top of the stack will be taken as the final product.
The product molecule is considered to be a \textit{projection} of the input molecular graph in the chemical space. 
We select the final product molecules according to properties of interest, such as similarity and docking scores.

\subsection{Experiment Setup}

\paragraph{Reaction templates}
We use the SynNet reaction template set \cite{gao2021synnet} curated by the authors from two publicly available template sets \cite{hartenfeller2012dogs,button2019automated}.
The template set contains 91 reaction templates encoded in SMARTS string format \citep{DaylightTheory2023}, of which 13 are uni-molecular reactions and 78 are bi-molecular reactions.

\paragraph{Building blocks}
We use the building blocks in the Enamine US Stock catalog retrieved in October 2023 \cite{enamine}.
For building blocks that include more than one molecule (\textit{e.g.}, salts, hydrates), we keep the largest molecule and drop the remaining ones.
Building blocks that fail the RDKit sanitization check or do not match any reaction template are removed.
Duplicate building blocks are also removed.
The pre-processing procedure leads to the final building block set containing 211,220 molecules.

\paragraph{Similarity scores}
To evaluate the similarity between the input molecule and the output molecule, we use the Tanimoto similarity score on three different fingerprints: (1) Morgan fingerprint of length 4096 and radius 2 \citep{morgan1965generation}, (2) Morgan fingerprint of Murcko scaffold, and (3) Gobbi pharmacophore fingerprint \citep{gobbi1998genetic}.
The three similarity scores indicate chemical similarities in three different aspects: overall structure, scaffold structure, and pharmacophore property, respectively, and are all normalized to $[0, 1]$.

\paragraph{Code and data} The code and data of this project are available at \url{https://github.com/luost26/ChemProjector}.
\section{Results}

\begin{figure*}[ht]
\begin{center}
\centerline{\includegraphics[width=1.0\linewidth]{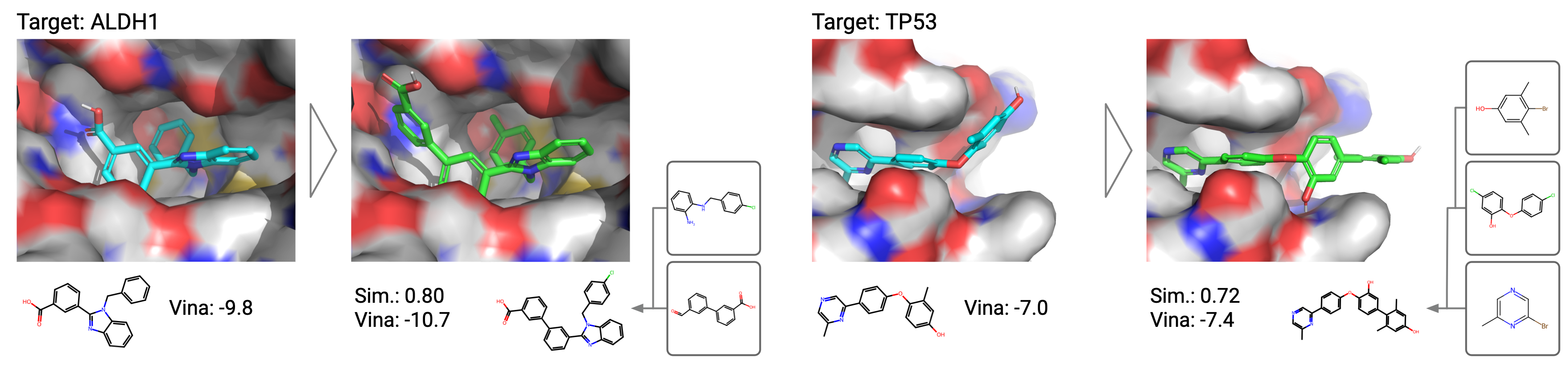}}
\vspace{-10pt}
\caption{Examples of projecting molecules generated by Pocket2Mol to synthesizable analogs. The projection improves the Vina score and proposes synthesis pathways. The docking poses remain similar.}
\label{fig:docking}
\end{center}
\vspace{-25pt}
\end{figure*}

In this section, we first examine the model's capability of finding synthesis pathways for molecules that are known to be synthesizable, and explore the generalization ability of the models (Section~\ref{sec:results:synthesis}).
Then, we study the application of the model in three drug design settings: structure-based drug design (Section~\ref{sec:results:sbdd}), goal-directed generation (Section~\ref{sec:results:goal}), and hit expansion (Section~\ref{sec:results:hit_expansion}).

\subsection{Generalization in Bottom-Up Synthesis Planning}
\label{sec:results:synthesis}
We first validate our model's ability in bottom-up synthesis planning, particularly in proposing synthetic pathways for novel molecular structures.
To assess this capability with unseen molecules, we employed the following testing approach.
First, we clustered building blocks into 128 groups using K-means algorithm based on their fingerprints, reserving one cluster exclusively for testing while using the remaining 127 for training.
This approach generated a test set of 1,000 molecules from the reserved building block to evaluate our model's generalization.
During testing, we allow the model to access the reserved building blocks.
In addition, we include another challenging benchmark: 1,000 molecules from the ChEMBL database (Release 33)\citep{gaulton2012chembl}, reported to be predominantly ``unreachable'' in previous work \citep{gao2021synnet}.
This not only tested our model against complex, unseen molecular structures but also set a strenuous standard for assessing its generalizability.

We evaluate the model using the following metrics:
(1) success rate: the percentage of valid postfix notations;
(2) reconstruction rate: the percentage of proposed synthetic pathways that lead to products identical to input molecules;
(3) average similarities. The similarity score is set to 0 for failed cases.
We compare our model with the baseline model SynNet \citep{gao2021synnet}.

Our model produces valid synthetic pathways for most test cases and thus achieves a high success rate (Table~\ref{tab:generalize}).
Most notably, 29.3\% of the test molecules and 13.4\% of the ChEMBL molecules have been successfully reconstructed. Furthermore, the similarity scores of the proposed model are significantly higher than the baseline model.
The result demonstrates that our model exhibits generalizability to unseen molecules.

\begin{table}[t]
\vspace{-10pt}
  \caption{Similarity scores between molecules generated by Pocket2Mol and their analogs.}
  \label{tab:sbdd-sim}
  \vskip 0.1in
  \centering
  \small
  \begin{adjustbox}{max width=1\columnwidth}
  \begin{tabular}{l ccc}
\toprule
 & Sim. & Sim. & Sim.  \\
 & (Morgan) & (Scaffold) & (Gobbi) \\
\midrule
ADRB2 & 0.5804 & 0.6362 & 0.4409 \\
ALDH1 & 0.3875 & 0.3399 & 0.3034 \\
ESR1\_ago & 0.4066 & 0.3520 & 0.2990 \\
ESR1\_ant & 0.4965 & 0.4919 & 0.4122 \\
FEN1 & 0.4180 & 0.4090 & 0.3369 \\
GBA & 0.3140 & 0.2572 & 0.2460 \\
IDH1 & 0.3830 & 0.3423 & 0.3178 \\
KAT2A & 0.5028 & 0.4831 & 0.4378 \\
MAPK1 & 0.4274 & 0.3942 & 0.3886 \\
MTORC1 & 0.4565 & 0.4193 & 0.3820 \\
OPRK1 & 0.5125 & 0.5428 & 0.4312 \\
PKM2 & 0.4558 & 0.4315 & 0.3818 \\
PPARG & 0.4992 & 0.4884 & 0.4416 \\
TP53 & 0.5566 & 0.5386 & 0.4617 \\
VDR & 0.4355 & 0.3857 & 0.3615 \\
\bottomrule
\end{tabular}

  \end{adjustbox}
\end{table}

\begin{table}[t]
\vspace{-10pt}
  \caption{The analogs slightly improve Vina scores, suggesting the binding is likely to be preserved.}
  \label{tab:sbdd-vina}
  \vskip 0.1in
  \centering
  \small
  \begin{adjustbox}{max width=1\columnwidth}
  \begin{tabular}{l cccc}
\toprule
 & \multicolumn{4}{c}{Vina (kcal/mol)}
\\
Target & Ref. & Gen. & Analog ($\downarrow$) & $\Delta$ ($\downarrow$) \\
\midrule
ADRB2 & -8.24 & -8.70 & -9.09 &  -0.38 \\
ALDH1 & -8.49 & -9.10 & -9.00 & 0.10 \\
ESR1\_ago & -8.28 & -9.46 & -9.36 & 0.09 \\
ESR1\_ant & -7.62 & -9.77 & -9.77 & 0.00 \\
FEN1 & -6.32 & -6.54 & -6.60 &  -0.06 \\
GBA & -8.14 & -6.96 & -7.69 &  -0.73 \\
IDH1 & -9.19 & -9.50 & -9.56 &  -0.07 \\
KAT2A & -7.18 & -8.07 & -8.18 &  -0.11 \\
MAPK1 & -8.63 & -8.61 & -8.98 &  -0.37 \\
MTORC1 & -9.64 & -10.17 & -10.20 &  -0.04 \\
OPRK1 & -8.83 & -8.36 & -8.77 &  -0.40 \\
PKM2 & -9.23 & -9.45 & -10.09 &  -0.65 \\
PPARG & -7.16 & -8.47 & -8.53 &  -0.06 \\
TP53 & -6.13 & -7.11 & -7.47 &  -0.36 \\
VDR & -8.81 & -10.51 & -10.46 & 0.05 \\
\bottomrule
\end{tabular}

  \end{adjustbox}
\end{table}

\begin{table*}[tb]
\vspace{-10pt}
  \caption{In general, the molecules generated by goal-directed generative models do not show significant similarity to their analogs. The projection typically results in a drop in the objective scores. Nonetheless, the maximum scores of the analogs on 6 targets decrease by an amount of less than 0.1, indicating that the properties are still preserved for some molecules.}
  \label{tab:guacamol}
  \vskip 0.1in
  \centering
  \small
  \begin{adjustbox}{max width=1\textwidth}
  \begin{tabular}{lccccccccc}
\toprule
 & Sim. & Sim. & Sim. & \multicolumn{3}{c}{Avg(Objective)} & \multicolumn{3}{c}{Max(Objective)} \\
Property & Morgan & Scaffold & Gobbi & Gen. & Analog & $\Delta$ & Gen. & Analog & $\Delta$ \\
\midrule
Amlodipine MPO & 0.4593 & 0.3098 & 0.3855 & 0.8110 & 0.4824 & -0.3286 & 0.8895 & 0.8793 & -0.0102 \\
Deco Hop & 0.5499 & 0.8273 & 0.7173 & 0.9768 & 0.7296 & -0.2472 & 0.9992 & 0.9652 & -0.0340 \\
Fexofenadine MPO & 0.4510 & 0.4589 & 0.3431 & 0.9316 & 0.4321 & -0.4995 & 1.0000 & 0.8999 & -0.1001 \\
Osimertinib MPO & 0.4339 & 0.3956 & 0.5620 & 0.9030 & 0.5424 & -0.3606 & 0.9508 & 0.9326 & -0.0182 \\
Perindopril MPO & 0.3175 & 0.2605 & 0.3181 & 0.7169 & 0.3552 & -0.3617 & 0.7850 & 0.7011 & -0.0839 \\
Ranolazine MPO & 0.4145 & 0.4648 & 0.5029 & 0.8889 & 0.5652 & -0.3237 & 0.9103 & 0.8377 & -0.0726 \\
Scaffold Hop & 0.5092 & 0.7499 & 0.6125 & 0.9558 & 0.5283 & -0.4275 & 1.0000 & 0.8993 & -0.1007 \\
Sitagliptin MPO & 0.2536 & 0.2230 & 0.2427 & 0.6527 & 0.0158 & -0.6369 & 0.8437 & 0.4909 & -0.3528 \\
Valsartan SMARTS & 0.3500 & 0.3558 & 0.2572 & 0.8191 & 0.0036 & -0.8155 & 0.9861 & 0.4776 & -0.5085 \\
Zaleplon MPO & 0.5328 & 0.5653 & 0.5429 & 0.6379 & 0.2502 & -0.3877 & 0.7151 & 0.7122 & -0.0029 \\
\bottomrule
\end{tabular}

  \end{adjustbox}
\end{table*}

\subsection{Projecting Molecules Generated by Structure-Based Drug Design Models}
\label{sec:results:sbdd}

Generative models for structure-based drug design (SBDD) build molecules to fit a protein pocket structure \citep{luo20213sbdd}.
SBDD models often produce chemically invalid structures, because they operate in the 3D space with a few chemical constraints \citep{harris2023benchmarksbdd}. 
As a remedy, we apply our model to molecules proposed by the SBDD models to find valid and synthesizable analogs.

We use Pocket2Mol \cite{peng2022pocket2mol} to generate molecules for targets in the LIT-PCBA dataset \cite{tran2020litpcba}.
The LIT-PCBA dataset contains 15 protein targets, each of which has at least one X-ray structure that can be used to run Pocket2Mol to generate molecules.
For each target, we select 300 generated molecules based on the sum of QED and SA scores.
Then, we apply our model to generate 5 analogs for each molecule and select best one according to Vina scores \citep{eberhardt2021autodock}. 
The estimated binding energy in kcal/mol shows slight improvement for the majority of targets (Table~\ref{tab:sbdd-vina}), suggesting that the top-1 synthesizable analogs of each molecule could maintain binding.
Figure~\ref{fig:docking} presents two examples in which the synthesizable analogs improve predicted binding. Figure~\ref{fig:sbdd-samples-1} and \ref{fig:sbdd-samples-2} in the appendix present more examples.

\subsection{Projecting Molecules Generated by Goal-Directed Generative Models}
\label{sec:results:goal}

Goal-directed generative models are pivotal in molecular design, yet their practicality is often hampered by the synthesizability of the produced molecules.
This issue was highlighted by \citeauthor{gao2020synthesizability}, who, utilizing three algorithms on the GuacaMol benchmark suite's multi-property objectives \citep{brown2019guacamol}, revealed that around 70\% of the molecules generated were considered unsynthesizable by the ASKCOS retrosynthesis analysis tool \citep{coley2019askcos}.

We applied our model to these molecules which were labeled as ``unsynthesizable'' by ASKCOS.
The analogs show limited structural similarity with an average Morgan fingerprint similarity score of 0.43 (Table~\ref{tab:guacamol}).
This divergence in structure is primarily due to the non-synthesizability of the original molecules, making it hard to compose a similar analog using the provided building blocks and reactions of the chemical space.
The property scores achieved by the original unsynthesizable molecules are affected by the structural discrepancies, as indicated by the drop in average objective score.
However, for most of the targets, there still exists at least one analog that mostly retains the original property as shown by the maximum objective scores.

In conclusion, the projection generally leads to synthesizable analogs that are sub-optimal in terms of the oracle property scores.
These sub-optimal analogs share some common structural factors with the original molecules, hopefully making them good starting points for combinatorial optimization within the chemical space \citep{gao2021synnet,swanson2023generative}.

\begin{figure}[t]
\begin{center}
\centerline{\includegraphics[width=1.0\columnwidth]{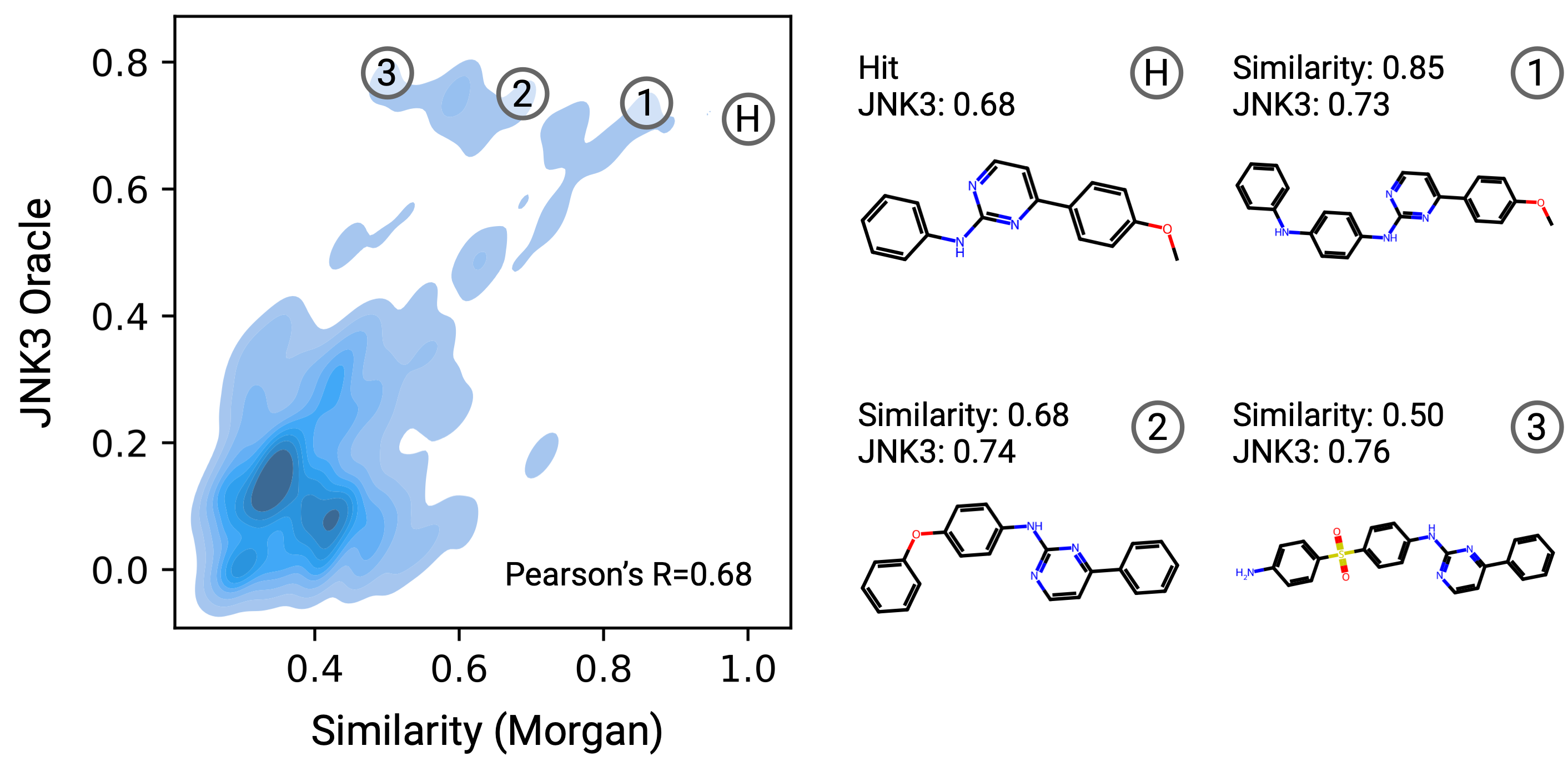}}
\vspace{-10pt}
\caption{The distribution of JNK3 oracle scores and similarity of analogs generated by expanding the hit molecule.}
\label{fig:hitexp}
\end{center}
\vspace{-15pt}
\end{figure}

\subsection{Projecting Molecules to Explore Local Chemical Space for Hit Expansion}
\label{sec:results:hit_expansion}

Our model's capabilities also extend to hit expansion in drug discovery, offering a strategic approach to identifying structurally similar, synthesizable analogs of hit molecules \cite{keserHu2006hit,levin2023hitexpand}.

We showcase our model's hit expansion capability by designing inhibitors for c-Jun N-terminal Kinases-3 (JNK3), following the setting of \citeauthor{levin2023hitexpand}. Based on the oracle function provided by \citeauthor{li2018multi}, our goal is to identify high-scoring, synthesizable molecules similar to a specified hit molecule with a JNK3 score of 0.68.
We generated 500 analogs for the hit, resulting in a diverse array of analogs with different JNK3 scores and structural similarities, as depicted in Figure~\ref{fig:hitexp}. Remarkably, 21 analogs outperformed the original hit in JNK3 score while maintaining an average Tanimoto similarity of 0.67.
Additionally, these analogs present distinct synthetic pathways, different from traditional synthesis planning and building block exchange strategies \citep{levin2023hitexpand}, underlining our model's capability to cover broader chemical space (Figure~\ref{fig:jnk3-synth} in the appendix).

\section{Conclusion}

In this work, we introduce a method that projects molecules proposed by generative models into synthesizable chemical space.
The method is composed of a novel postfix notation-based representation of synthesis and a transformer-based encoder-decoder network.
We study the application of the method in structure-based drug design, goal-directed molecular generation, and hit expansion.
The results demonstrate that the model can effectively convert the generated molecules into synthesizable analogs while maintaining the desired properties.
As opposed to previous methods that rely on combinatorial optimization or virtual screening, the proposed approach enables efficient exploration of the chemical space by utilizing the flexibility of generative models while respecting synthetic feasibility, further unlocking the potential of machine learning in molecular design.

\newpage

\section*{Acknowledgement}

J.M. acknowledges the support through the National Key Research and Development Program of China grant 2022YFF1203100.
W.G. is supported by the Google Ph.D. Fellowship and Office of Naval Research under grant number N00014-21-1-2195. 
C.W.C. thanks the AI2050 program at Schmidt Futures (grant G-22-64475) for financial support.

\section*{Impact Statement}
This paper presents work whose goal is to advance the generative modeling of molecular structures. There are many potential societal consequences of our work, none of which we feel must be specifically highlighted here.

\bibliography{references}
\bibliographystyle{icml2024}

\newpage
\appendix
\onecolumn

\section{Implementation}

\subsection{Training Data}
The algorithm for the construction of (postfix notation, molecular graph) pairs is described below:

\begin{algorithm*}[h]
   \caption{Construct a (postfix notation, molecular graph) pair for training}
   \label{alg:train}
   \begin{algorithmic}
  \STATE {\bfseries Input:} $m_r$, maximum number of reactions
  \STATE {\bfseries Input:} $m_a$, maximum number of atoms
  \STATE {\bfseries Input:} $\gF: M \mapsto \{ R \}$, a function that returns available reactions for molecule $M$. When $M = \texttt{null}$, it returns all the reactions $\gR$.
  \STATE {\bfseries Global:} $\gH: R \mapsto \{ M \}$, a function that stores available building blocks for each reaction.
  \STATE {\bfseries Initialize: } $S \gets []$ (empty stack)
  \STATE {\bfseries Initialize: } $P \gets []$ (empty postfix notation)
  
  \WHILE{$\operatorname{CountReactions}(P) < m_r$ and $\operatorname{CountAtoms}( \operatorname{top}(S) ) < m_a$}
  
      \STATE {\bfseries Let} $R \sim \gR(\operatorname{top}(S))$ (randomly choose a reaction)
      \STATE {\bfseries Let} $\{ B_i : 1 \le i \le (\operatorname{NumReactants}(R) - \operatorname{len}(S)) \} \sim \gH(R)$ (randomly choose reactants for $R$)
    
      \FOR{$1 \le i \le (\operatorname{NumReactants}(R) - \operatorname{len}(S))$}
        \STATE {\bfseries Push} $B_i$ onto $S$
        \STATE {\bfseries Append} $B_i$ to $P$
      \ENDFOR
      \STATE {\bfseries Pop} $\operatorname{NumReactants}(R)$ elements from $S$ to {\bfseries Call} $R$; Then {\bfseries Push} the product $Y$ onto $S$.
      \STATE {\bfseries Append} $R$ to $P$

  \ENDWHILE
  \STATE {\bfseries Return} $P$, $\operatorname{top}(S)$ (postfix notation and its product molecular graph)

\end{algorithmic}
\end{algorithm*}

\subsection{Parameters}
The molecular graph encoder consists of 8 graph transformer layers, each of which has 8 attention heads, and the dimension of the input and output features is 512.
The postfix notation decoder is a stack of 8 transformer decoder layers. Each has 8 attention heads, and the dimension of the input and output features is also 512.
In the nearest-neighbor search, each building block molecule is indexed by the Morgan fingerprint \cite{morgan1965generation} of length 256 and radius 2.
We use the AdamW optimizer \cite{loshchilov2017adamw} to train the network with a learning rate of 3e-4 and a batch size of 256 for 500,000 iterations.
During the construction of synthesis for training, we limit the maximum number of reactions to 5 and the maximum number of atoms to 80.
At inference time, we set the maximum sequence length to 16. If the end token is not generated within the limit, the process will be considered unsuccessful.

\section{Additional Results}

\subsection{Stricter Test Split for Bottom-Up Synthesis Planning}

To test the model's generalization more strictly, we construct a refined subset of the test set described in Section~\ref{sec:results:synthesis}.
Specifically, for each building block in the test set, we identify its most similar counterpart in the training set and calculate the maximum training set similarity accordingly.
Building blocks with a maximum training set similarity greater than 0.6 are then removed, resulting in a refined set of test building blocks.
We use this refined building block set to generate 1,000 molecules to test our model and SynNet, and the results are presented below.

\begin{table*}[h]
  \centering
  \small
  \begin{adjustbox}{max width=1\textwidth}
  \begin{tabular}{l ccccc}
\toprule
Method & Success\% & Recons.\% & Sim.(Morgan) & Sim.(Scaffold) & Sim.(Gobbi) \\
\midrule
SynNet & 84.0\% & 10.0\% & 0.4552 & 0.6012 & 0.3348 \\
\bf ChemProjector & \bf 95.8\% & \bf 17.0\% & \bf 0.5921 & \bf 0.7239 & \bf 0.6110 \\
\bottomrule
\end{tabular}

  \end{adjustbox}
\end{table*}

Despite the stricter partition affecting the performance of both models, our model demonstrates superior generalization. Particularly noteworthy is the higher scaffold similarity, indicating that our model better preserves the scaffold of unseen molecules.

\subsection{Baseline for Goal-Directed Generation}

We present below a comparison between our model and the baseline model SynNet in the goal-directed generation setting.
Our model outperforms the baseline in 7 out of 10 objectives in both similarity and property scores. Additionally, our model's performance is on par with the baseline in the remaining 3 objectives.
\begin{table*}[h]
  \centering
  \small
  \begin{adjustbox}{max width=1\textwidth}
  \begin{tabular}{l cc cc cc cc cc}
\toprule
 & \multicolumn{2}{c}{Sim. Morgan} 
 & \multicolumn{2}{c}{Sim. Scaffold}  
 & \multicolumn{2}{c}{Sim. Gobbi} 
 & \multicolumn{2}{c}{Avg(Obj)}
 & \multicolumn{2}{c}{Max(Obj)} \\
Property & SynNet & Ours & SynNet & Ours & SynNet & Ours & SynNet & Ours & SynNet & Ours \\
\midrule
Amlodipine MPO & 0.3732 & \textbf{0.4593} & 0.2333 & \textbf{0.3098} & 0.2787 & \textbf{0.3855} & 0.3358 & \textbf{0.4824} & 0.8605 & \textbf{0.8793} \\
Deco Hop & 0.4854 & \textbf{0.5499} & 0.5385 & \textbf{0.8273} & 0.5535 & \textbf{0.7173} & 0.6434 & \textbf{0.7296} & 0.9474 & \textbf{0.9652} \\
Fexofenadine MPO & 0.4308 & \textbf{0.4510} & 0.3942 & \textbf{0.4589} & 0.2983 & \textbf{0.3431} & 0.3501 & \textbf{0.4321} & 0.7928 & \textbf{0.8999} \\
Osimertinib MPO & 0.2375 & \textbf{0.4339} & 0.1740 & \textbf{0.3956} & 0.2494 & \textbf{0.5620} & 0.4827 & \textbf{0.5424} & 0.8613 & \textbf{0.9326} \\
Perindopril MPO & \textbf{0.3184} & 0.3175 & 0.2402 & \textbf{0.2605} & 0.2900 & \textbf{0.3181} & 0.2214 & \textbf{0.3552} & 0.6789 & \textbf{0.7011} \\
Ranolazine MPO & 0.3528 & \textbf{0.4145} & 0.2972 & \textbf{0.4648} & 0.2895 & \textbf{0.5029} & 0.3628 & \textbf{0.5652} & \textbf{0.8451} & 0.8377 \\
Scaffold Hop & 0.3694 & \textbf{0.5092} & 0.4965 & \textbf{0.7499} & 0.5082 & \textbf{0.6125} & 0.4593 & \textbf{0.5283} & 0.5299 & \textbf{0.8993} \\
Sitagliptin MPO & \textbf{0.3309} & 0.2536 & \textbf{0.2852} & 0.2230 & \textbf{0.2711} & 0.2427 & \textbf{0.0302} & 0.0158 & 0.4648 & \textbf{0.4909} \\
Valsartan SMARTS & \textbf{0.3742} & 0.3500 & \textbf{0.3805} & 0.3558 & \textbf{0.2771} & 0.2572 & \textbf{0.0074} & 0.0036 & \textbf{0.6034} & 0.4776 \\
Zaleplon MPO & 0.5141 & \textbf{0.5328} & 0.4884 & \textbf{0.5653} & 0.4856 & \textbf{0.5429} & 0.1477 & \textbf{0.2502} & 0.5548 & \textbf{0.7122} \\
\bottomrule
\end{tabular}

  \end{adjustbox}
\end{table*}

\subsection{Baseline for Hit Expansion}

In hit expansion, the objective is to identify similar molecules within the local chemical space.
Given that random molecular edits may not guarantee synthesizability, we adopt querying a chemical database as the baseline method.
Specifically, we search PubChem \citep{kim2023pubchem} using the hit molecule and retrieve the most similar 500 molecules based on fingerprint similarity. Only 7 out of the retrieved molecules surpass the hit in JNK3 score, with an average similarity score of 0.67. This result underscores the efficiency of our model in identifying relevant synthesizable compounds within the chemical space.

\newpage

\subsection{Molecules Generated by Pocket2Mol and Analogs}

\begin{figure*}[h!]
\begin{center}
\centerline{\includegraphics[width=\linewidth]{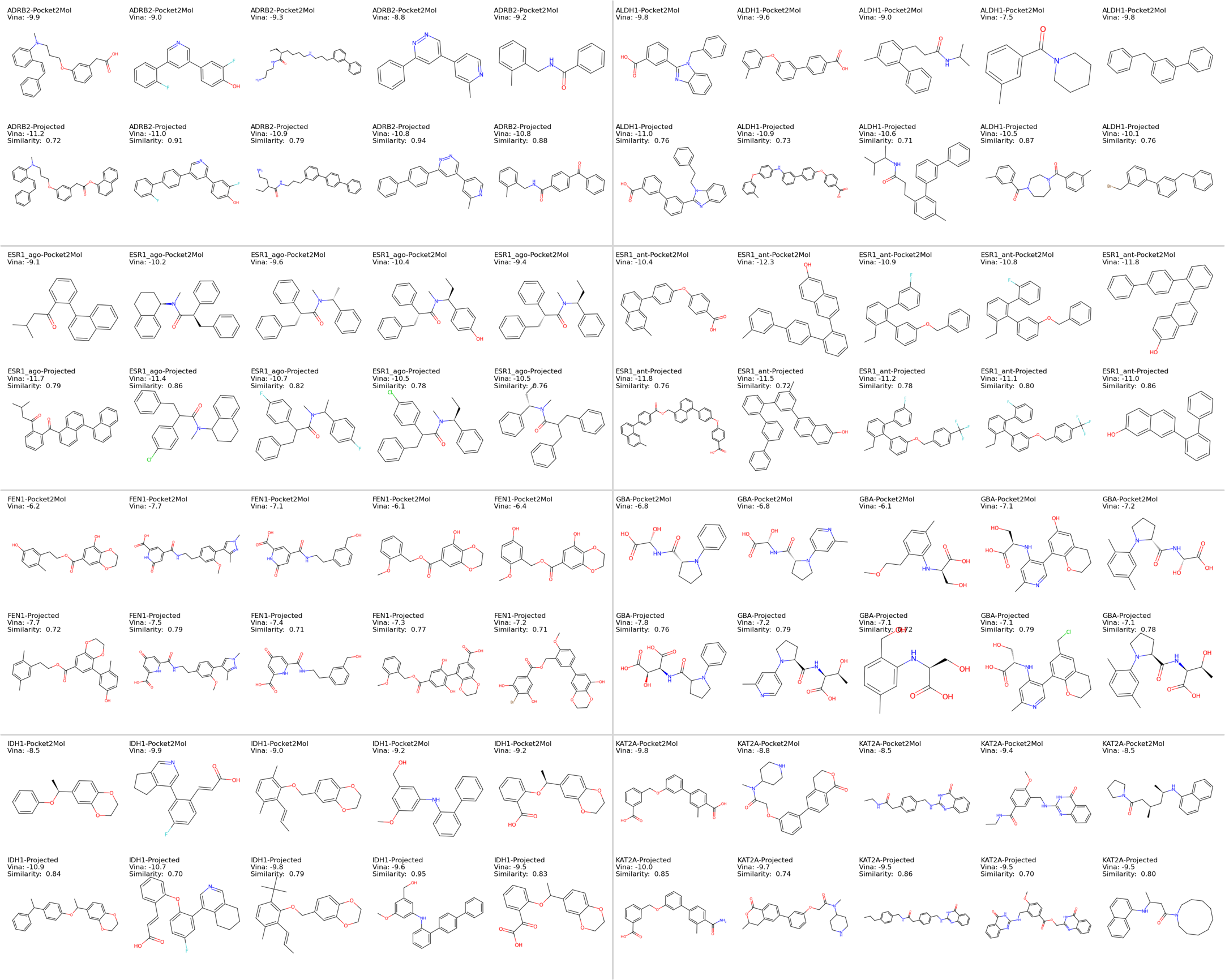}}
\caption{Molecules generated by Pocket2Mol and their analogs (1/2).}
\label{fig:sbdd-samples-1}
\end{center}
\end{figure*}

\newpage

\begin{figure*}[h!]
\begin{center}
\centerline{\includegraphics[width=\linewidth]{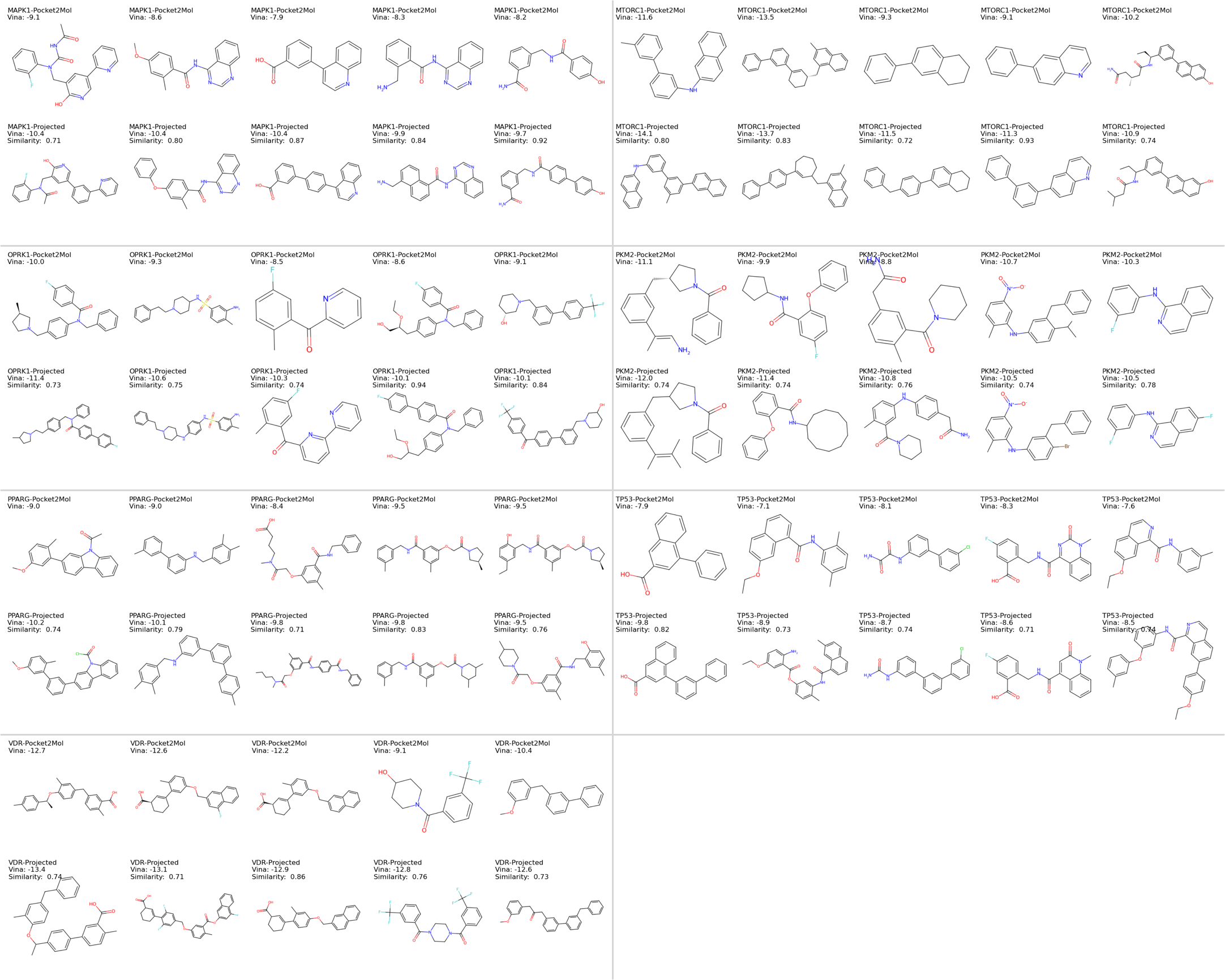}}
\caption{Molecules generated by Pocket2Mol and their analogs (2/2).}
\label{fig:sbdd-samples-2}
\end{center}
\end{figure*}

\newpage

\subsection{Synthetic Pathways in Hit Expansion}

\begin{figure*}[h!]
\begin{center}
\centerline{\includegraphics[width=0.9\linewidth]{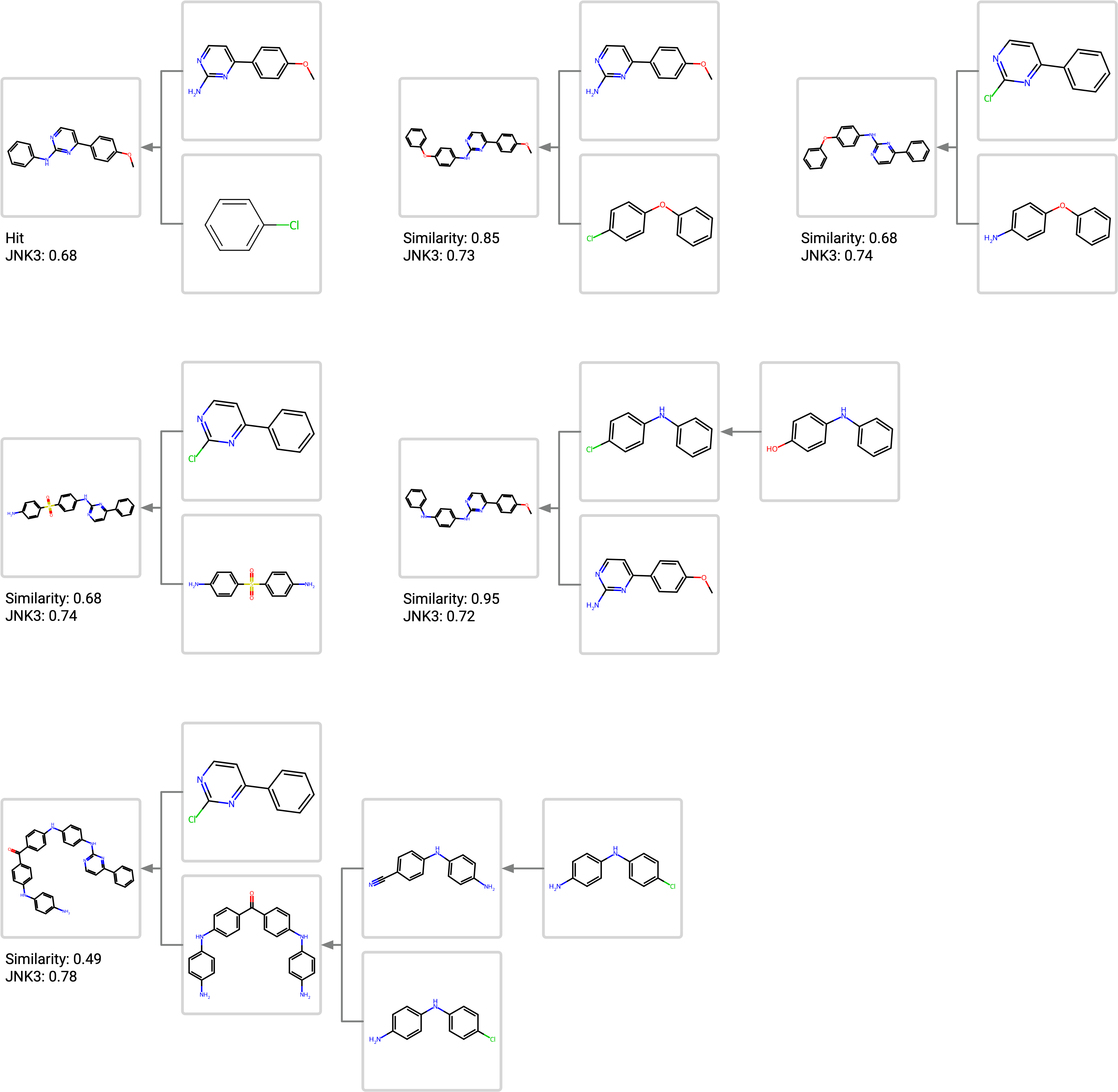}}
\caption{Distinct synthetic pathways of the analogs in hit expansion.}
\label{fig:jnk3-synth}
\end{center}
\end{figure*}

\end{document}